# Automatic Organisation and Quality Analysis of User-Generated Content with Audio Fingerprinting

Gonçalo Mordido, João Magalhães, Sofia Cavaco
NOVA LINCS, Departamento de Informática
Faculdade de Ciências e Tecnologia,
Universidade Nova de Lisboa
2829-516 Caparica, Portugal
goncalomordido@gmail.com, {jmag, scavaco}@fct.unl.pt

*Abstract*—The increase of the quantity of user-generated content experienced in social media has boosted the importance of analysing and organising the content by its quality. Here, we propose a method that uses audio fingerprinting to organise and infer the quality of user-generated audio content. The proposed method detects the overlapping segments between different audio clips to organise and cluster the data according to events, and to infer the audio quality of the samples. A test setup with concert recordings manually crawled from YouTube is used to validate the presented method. The results show that the proposed method achieves better results than previous methods.

## I. INTRODUCTION

The abundance and ubiquity of user-generated content has increased the demand for tools for the organisation and analysis of vast and heterogeneous data. Most of the activity experienced in social networks today contains audio excerpts, either from video files or actual audio clips. Therefore, the analysis of the audio features present in such content can contribute with relevant information for managing the data and ultimately provide a better experience to the end-user.

We propose a method that uses audio features to organise and determine the quality of user-generated audio content crawled from social media websites. In particular, we focus on data related to concert clips. The existence of several recordings of a given event, creates an abundant and redundant pool of recordings. As such, musical shows represent a very good use case for the presented work. The proposed method shall act as a way to better understand and deal with extensive datasets of audio files. The inference of the quality of each file within a given group of files has the ultimate goal of promoting a better user experience, since only the high-quality clips should be shown to the end-user.

The proposed method detects the overlapping sections between different audio clips to organise and group (*i.e.* cluster) the data. It then infers the audio quality of the samples directly from the features used to perform the clustering. The method uses an audio fingerprinting algorithm to this end.

Audio fingerprinting algorithms have traditionally been used for music recognition as made famous by Shazam, where a query sample is matched against other samples in a database of audio files [1, 2, 7, 9]. Here we use this technique for a different purpose, more specifically to synchronise different samples and use the synchronisation information to perform their clustering and infer their quality. In fact, other authors have shown that audio fingerprinting can be used to perform the synchronisation between different samples from the same event that are not time aligned [3, 5, 6, 8].

While Kennedy and Naaman have also used audio fingerprinting to this end [6], we propose two important improvements: (1) our clustering phase includes a filtering approach to avoid false positives, and (2) the proposed technique to infer the samples quality uses information from the audio fingerprinting algorithm that was not used before. Consequently, the analysis of the ranking of the samples in terms of quality achieves better results with the proposed method than with previous methods, with the assumption that professional edited recordings should have higher quality scores than user-generated recordings.

## II. THE DATA ORGANISATION METHOD

The proposed method can be used to organise multiple concert recordings, which here we call samples. There may be several samples from the same music. More specifically, the method focuses on the grouping of the audio samples, based on them having a common segment of audio, and on their relative quality inference. Since these samples are generated from user recordings, some challenges need to be tackled such as those related to the audio recording qualities inherent to each device. Moreover, it is very unlikely that any two recordings are time synchronised and have the same duration.

In practical terms, the information retrieved by our method can be used to organise and aid with the choice of which overlapped recordings to use at a given time based on their quality (figure 1). Several steps must be followed to perform the grouping and quality analysis of the different audio samples.

[2]Image from: COGNITUS, http://cognitus-h2020.eu/index.php/2017/01/06/two-open-access-datasets-with-user-generated-audio-recordings/

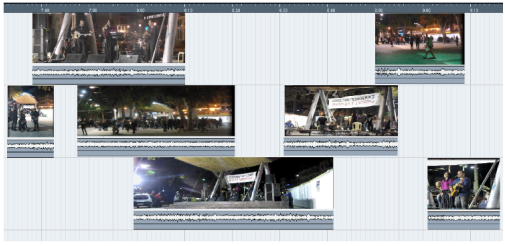

Fig. 1. Time-aligned audio clips[2]. Information about the quality of the clips is very useful to choose which clip should be played.

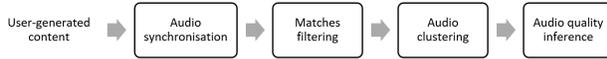

Fig. 2. Diagram of the different proposed steps.

To promote a better comparison between the different samples, their (1) synchronisation is required (for recordings of the same song or event). The synchronisation results will serve as the basis to perform the (2) grouping of the samples: recordings of the same song will be clustered together. The clustering information will ultimately help in deriving the (3) quality of each sample within a cluster. Evaluation and analysis of the attributed cluster and quality score of each sound clip must be also performed to validate the proposed system. Figure 2 illustrates how the different steps are executed in chronological order. These steps are described in detail in sections II-A, II-B, and II-C, with further validation being presented in section III.

*A. Audio Synchronisation*

The synchronisation of the samples is an essential step to perform the audio clustering, since its information is used to group different samples with common overlapping audio segments. Moreover, such information is also used to derive the quality of the different samples inside a given cluster.

Fingerprinting techniques' resistance to noise is particularly relevant when dealing with low quality music recordings. This characteristic is ideal for our method since it enables to synchronise noisy samples against possibly less noisy samples in the database, while only needing possibly short length common segments to synchronise different clips.

Fingerprint generation over any kind of data is an efficient mechanism to characterise possibly large data with a small representation. More explicitly, as an alternative to representations that involve large amounts of data, fingerprints are compact representations of the data that can be used for purposes that do not require dealing with all the intrinsic details of the data. This technique promotes a fast way of comparing the quality of two entities by trying to diminish the need to compare irrelevant features.

There are several steps commonly inherent to any audio fingerprint technique: first, an extraction of some of the features present in the audio clip is done; second, those features, or a combination of them, are used to generate a fingerprint that will characterise the audio clip; then, searching processes are applied to access the fingerprints of the other clips in the database; and finally matching mechanisms compare pairs of fingerprints and generate matches.

While other fingerprinting algorithms are compatible with our method, we used Cotton and Ellis fingerprinting algorithm Python implementation[3] based on the methodologies proposed by Wang [4, 9]. The fingerprint generation proposed in this approach is landmark-based (*i.e.* fingerprints are composed by several landmarks). A landmark is created by the analysis of frequency peaks with high energy, since these high spectral energy characteristics of the songs are likely to be resistant to noise and distortion [4]. A landmark is a pair of two peaks, and contains information about each peak frequency, the time at which the first peak occurred, and the time offset between them.

Two samples are considered to match if they have more than $t_l$ common landmarks. Since false matches are unlikely (i.e. have a low probability of occurrence but still greater than 0), this threshold can have a small value (such as $t_l = 5$, being this the default value in the audio fingerprinting algorithm used). For each sample $s$ in a database of previously added samples, the audio fingerprinting phase finds the **matching list** for sample $s$, that is, all samples that match $s$. This is done by taking into account the number of common landmarks between query sample $s$ and each of the other samples in the database. The common landmarks are called **matching landmarks**.

*B. Audio Clustering*

Since we are dealing with several concert recordings as the context of our problem, events can be characterised as the different songs played in the different concerts. Thus, the goal of the clustering phase is to group all recordings of a given concert's song in the same cluster by the analysis of common audio segments.

*1) Audio samples grouping:* As proposed by Kennedy and Naaman, the results from the audio fingerprinting algorithm can be used to cluster the samples in events [6]. The matched samples have some matching landmarks, which is an indication that, potentially, the samples have a common excerpt and thus are recordings of the same song. The clustering phase uses this information to cluster together the samples that are matched in the audio fingerprinting phase. Therefore, we consider all database matches retrieved by the audio fingerprinting algorithm (*i.e.* all samples $s_i, \ldots, s_j$ matching a given query sample $s$).

To find the clusters, we represent the matches between different samples with a graph, $G$ [6]. Each sample in the database is represented by a vertex in $G$. The edges represent the matches between samples. In other words, if sample $s_1$ matches sample $s_2$, then there is an edge between vertex $s_1$ and vertex $s_2$[4]. The edge weight is the offset (in seconds) between the two samples. The whole graph can have several components. Each component of $G$ corresponds to a different cluster. If there is a path between two vertices, then the corresponding samples are in the same cluster. Isolated vertices represent unmatched samples, for which the algorithm could not find any match in the database.

---
[3]https://github.com/dpwe/audfprint

Even though unlikely, the probability of false positives samples retrieved by the algorithm is greater than zero, leading to the merging clusters that should not be merged. For example, if sample $s_1$ from song 1 and sample $s_2$ from song 2 are incorrectly matched, their clusters will be wrongly merged. In order to overcome this drawback, we introduced a filtering stage to the clustering algorithm (section II-B2). This filtering approach aims to optimise the clustering results.

*2) Matches filtering:* Considering all the matches retrieved by the audio fingerprinting algorithm would be ideal if false positives did not occur. These false positives can be **sample-level** or **landmark-level**. The first case happens when samples not referring to recordings of the same song are matched. The latter case happens when several landmarks are matched with different offsets over just two samples. For example, samples $s_1$ and $s_2$ from the same song have $l_1$ matching landmarks with offset $o_1$, $l_2$ matching landmarks with offset $o_2$, etc. It is important to notice that only one of the retrieve offsets may be correct, since two samples can only have one offset. Thus, all the other offsets shall be considered false positives.

Landmark-level false positives are easily detected by the repetition of a sample in a query sample's matching list (output from the audio fingerprinting phase). To tackle this problem, only the match with higher number of matching landmarks is considered while any other sample repetitions in the list are eliminated. In the example above, the match considered is the match with offset $o_i$ such that $l_i = max(l_1, ..., l_k)$.

To handle sample-level false positives, it is important to understand in which context they appear because, even though unlikely, their probability is greater than zero. The analysis of such cases, showed that false positives have a lower number of matching landmarks than the true positives in the same matching list. This also applies when comparing the percentage, $p$, of matched landmarks in the overall number of landmarks of false and true positives. That is, $\forall_{s_t, s_f}\ p_t > p_f$, where $s_t$ is a true positive and $s_f$ is a false positive, $p_t$ and $p_f$ are the percentages of matching landmarks of samples $s_t$ and $s_f$, respectively, and $p$ is defined as follows

$$p_i = \frac{l_i}{t_i}, \qquad (1)$$

where $l_i$ is the number of matching landmarks between sample $s_i$ and query sample $s$, and $t_i$ is the total number of landmarks of sample $s_i$. It was also observed that when we consider the percentage of matching landmarks in decreasing order, the slope that leads to a false positive (*i.e.* to $p_f$) is steeper. Following this analysis, an appropriate filtering approach would be:

> For each sample $s$ in the database, consider all the samples $s_i$ that match $s$ as retrieved from the audio fingerprinting phase. Let us assume we have $n$ such samples.
> 1. For all those matching samples, consider the percentage of matching landmarks in decreasing order: $(p_1, p_2, \ldots, p_n)$, where $p_1 \geqslant p_2 \geqslant \ldots \geqslant p_n$.
> 2. Analyse the derivative on all consecutive pairs of points $(p_1, p_2, \ldots, p_n)$ in the graph of the percentage of matching landmarks.
> 3. Finally, consider as matches all the samples $s_1$ to $s_j$ up to a point where the derivative of this graph is higher than a certain value. In other words, stop considering matches as soon as the percentage of matching landmarks significantly drops.

In order to achieve this, one can use a threshold, $t_d$, and the graph's slope: $\Delta_i = p_{i+1} - p_i$. If $\Delta_j \leqslant t_d$ and $\Delta_i > t_d$ for $i \in \{1, \ldots, j-1\}$, then the algorithm considers that only the samples up to sample $s_j$ are matches to $s$. That is, only samples $s_1, s_2, \ldots, s_j$ are considered as matches. All remaining samples $s_{j+1}, \ldots, s_n$ are considered as false positives. Our algorithm is using $t_d = -0.07$, as a result of fine-tuning this parameter to exclude all false positives in our dataset described in sub-section III-A.

This would be a reasonable approach to follow if the difference between the percentage of matching landmarks between consecutive true positives would never decrease significantly. Yet, due to the variety in quality and duration of user-generated content, such situation can occur and would cause a high number of true positives to be discarded. Thus, the filtering approach has to take into consideration more parameters to better choose when to filter the returned matches. Since the probability of finding false matches is low and so is their percentage of matched landmarks, a sample should only be considered a false positive if its percentage of matched landmarks is lower than the average for all the retrieved matching samples. Thus, the algorithm's step 3 can be changed to:

3. Consider as matches (1) all the samples $s_1$ to $s_k$ such that $\forall_{1 \leqslant i \leqslant k}\ p_i \geqslant avg(p1, p_2, \ldots, p_n)$ and (2) all the samples $s_{k+1}$ to $s_j$ such that $\Delta_j \leqslant t_d$ and $\forall_{k+1 \leqslant i \leqslant j-1}\ \Delta_i > t_d$.

For an easier analysis of the filtering process, figure 3 displays an example of a matching list with 8 samples in form of a graph with the distribution of the percentage of matching landmarks over the total number of landmarks of each matched sample. In this example, there is only one false positive, which is sample 8. Therefore, ideally this should be the only discarded sample. As observed, the proposed approach achieves this by considering the low slope point (sample 7) under the average of the percentage of landmarks (marked with the dashed line) as the last accepted match, and discarding all samples after it (*i.e.* sample 8). While there are other low gradient points (marked with a diamond) and some have slopes lower than $t_d$, the samples at these positions are not discarded because they are above the average line.

---

[4]While samples and vertices are different entities, here there is a one-to-one relationship between them. Thus, to simplify the notation, we will use the same name to represent samples and vertices. For instance, $s_i$ can represent sample $s_i$ and vertex $s_i$. Also, to simplify the explanations in the paper, we may refer to the vertices in $G$ as samples.

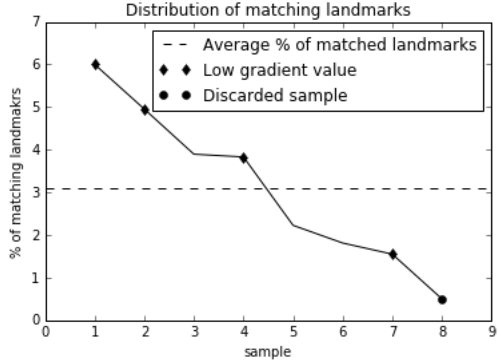

Fig. 3. Distribution of matching landmarks from the matching samples. The only false positive in the matching list (sample 8) is successfully discarded by our approach.

|  | #Instances | %Instances | #Filtered | %Filtered |
|---|---|---|---|---|
| Total matches | 620 | 100.00% | 70 | 11.29% |
| False positives (landmark-level) | 5 | 0.81% | 5 | 100.00% |
| False positives (sample-level) | 4 | 4.40% | 4 | 100.00% |
| False negatives (landmark-level) | - | - | 65 | 10.48% |

TABLE I
FILTERING BENCHMARKS.

*C. Audio Quality Inference*

The output of the clustering phase consists of a set of clusters, represented by graph $G$, that organises the data according to events (samples with a common segment of audio belong to the same cluster). The vertices within a component of $G$ may have a different number of neighbours as not all pairs of vertices within a component are adjacent. For instance, let $s_1$, $s_2$ and $s_3$ be vertices in $G$. If $s_1$ and $s_2$ are adjacent, and $s_2$ and $s_3$ are adjacent, all three vertices belong to the same component of $G$ (same cluster) but this does not mean that there is an edge between $s_1$ and $s_3$: $s_1$ and $s_2$ can have a common excerpt, $s_2$ and $s_3$ can have another common excerpt, and $s_1$ and $s_3$ may lack common excerpts.

Kennedy and Naaman propose to derive the quality of samples by the direct analysis of $G$, where a sample's quality is proportional to the number of neighbours of that sample's vertex, that is, the number of adjacent vertices [6]. In the example above, $s_1$ and $s_2$ are neighbours, $s_2$ and $s_3$ are neighbours but $s_1$ and $s_3$ are not neighbours. Samples with more neighbours, that is, samples that got matched the most by the fingerprinting algorithm, are considered as those with better quality. This is supported by the idea that any two low-quality samples are less likely to match each other than when at least one of the samples has good quality.

Even though this is a suitable approach, it only considers the number of matches of a given sample, ignoring how the sample ranks in terms of matching landmarks in the overall list of matched samples. A sample with many matching landmarks is likely to have better quality than a sample with less matching landmarks. Thus, our proposed solution considers that the score of a sample, $s_i$, depends on the total number of matching landmarks of that given sample against all the other samples in the database. Let $s_1, s_2, \ldots, s_N$ be the samples in the database. Assume $s_i$ (for $1 \leqslant i \leqslant N$) has $l_{1i}$ matching landmarks to sample $s_1$, $l_{2i}$ matching landmarks to sample $s_2$, etc. Then the score of $s_i$ is calculated as follows:

$$score(s_i) = \sum_{j=1}^{N} l_{ji}. \quad (2)$$

III. EVALUATION AND ANALYSIS

*A. Test setup*

To test the proposed algorithm, a realistic testbed was designed by manually collecting from YouTube several concert clips of various songs captured with different devices. For each song we collected one professional recording and several user recordings of the same song in the same concert.

The dataset[5] consists of 91 samples of 10 different songs, all part of different editions of the Reading Festival. Apart from the 10 professional recorded samples, all samples were recorded by users, which means different recording devices with different qualities apply. Moreover, the number of recordings retrieved for each song, as well as their respective lengths, differ between each cluster. Efforts were made to have recordings with lengths across a large scope in order to promote a more diverse dataset. Nonetheless most of the recordings crawled were within the 5 minutes range.

*B. Results*

Using the dataset described in III-A, we validated the proposed method, namely the filtering and clustering phases. By analysis of table I, we can see that our filtering method succeeded in filtering all false positives retrieved by the audio fingerprinting algorithm – 1 out of the 5 landmark-level false positives were discarded by eliminating repetitions whilst the rest were detected by the analysis of the derivative values.

It is important to notice that even though there were occurrences of false negatives (*i.e.* 10.48 % of the overall retrieved landmarks), this does not affect the clustering results, as presented in table II. The number of retrieved clusters when considering all matches and performing filtering is correct (that is, it is the same as for the ground-truth), whereas without the filtering process clusters were wrongly merged, making the system only retrieve 6 clusters instead of 10.

To evaluate the quality inference of our solution, we analysed the score of the professional recording relative to the scores of the other samples in the cluster using our method and Kennedy and Naaman's method (K.M. method). This evaluation assumes that the professional recordings represents one of the highest quality recordings in any cluster, and, therefore, should be placed at the top of the ranking list. Table III shows the comparison between the two methods. The second column indicates the number of samples in the cluster, the third and fourth columns show the position of the professional recording in the ranking list with our method and the K.M. method, respectively. The first position in the ranking list corresponds to the highest score.

---
[5]The dataset is available at http://novasearch.org/datasets/

|  | # Clusters | # Unmatched samples |
|---|---|---|
| Ground-truth | 10 | 0 |
| All matches (no filtering) | 6 | 2 |
| All matches (filtering) | 10 | 2 |

TABLE II
CLUSTERING BENCHMARKS.

| Cluster | # Samples | Proposed method | K.M. method |
|---|---|---|---|
| 1 | 8 | 5th | 3th-5th |
| 2 | 8 | 4th | 1st-4th |
| 3 | 6 | 5th | 3th-5th |
| 4 | 15 | 3th | 3th-4th |
| 5 | 11 | 1st | 2nd-4th |
| 6 | 5 | 4th | 1st-5th |
| 7 | 8 | 4th | 1st-4th |
| 8 | 10 | 1st | 2nd-4th |
| 9 | 11 | 5th | 3th-5th |
| 10 | 7 | 1st | 1st-4th |

TABLE III
QUALITY INFERENCE. POSITION OF THE PROFESSIONAL RECORDING IN THE RANKING LIST WITH OUR METHOD (THIRD COLUMN) AND KENNEDY AND NAAMAN'S METHOD (FOURTH COLUMN).

It is important to note that there may be ambiguity on the quality scores of the K.M. method, because several samples can have the same score. This is observed in table III by the range of positions of the professional sample. Such ambiguity does not appear with our approach, which is essential for aiding end users on the decision of which samples to use. The table also shows that the score of the professional track with our method is always the same or better than with the K.M method, as it is either included within the range given by the K.M method or it is even ranked better (clusters 5 and 8).

## IV. CONCLUSION

For a better comprehension and management of large datasets of audio files, we propose a method that clusters the data according to events and infers the relative quality of audio files. The method uses audio fingerprints to determine the clusters and quality of the samples. While our method uses some of the methodology proposed by Kennedy and Naaman [6], we propose relevant improvements to their methodology.

A major improvement offered by our method consists of avoiding false positives. On top of eliminating repetitions, our method uses a filtering approach that looks at the distribution of the percentage of matching landmarks (sets of fingerprints common to two samples) and uses the derivative of this distribution to detect false positives.

In terms of quality inference, by looking at the number of matching landmarks, instead of only checking if there was a match between two samples, the proposed method succeeds in classifying better the professional recordings in some cases to when compared to the method proposed in [6]. Furthermore, it also avoids ambiguity quality scores using more detailed information than the previous method.

The results show that the proposed filtering technique successfully avoids false negatives. Also, the quality inference results from our method show improvements over previous methods.

## V. ACKNOWLEDGEMENTS

This work was partially funded by the H2020 ICT project COGNITUS with the grant agreement No 687605 and by the project NOVA LINCS Ref. UID/CEC/04516/2013.